# Real Effect or Bias? Best Practices for Evaluating the Robustness of Real-World Evidence through Quantitative Sensitivity Analysis for Unmeasured Confounding

Authors: Douglas Faries[1], Chenyin Gao[2], Xiang Zhang[3], Chad Hazlett[4], James Stamey[5], Shu Yang[2], Peng Ding[6], Mingyang Shan[1], Kristin Sheffield[7], Nancy Dreyer[8].

Affiliations: 1 Real-World Access and Analytics, Eli Lilly & Company; 2 Department of Statistics, North Carolina State University; 3 Medical Affairs Biostatistics, CSL Behring; 4 Departments of Statistics and Political Science, University of California at Los Angeles; 5 Department of Statistical Science, Baylor University; 6 Department of Statistics, University of California Berkeley; 7 Value, Economics, and Outcomes, Eli Lilly & Company; 8 Dreyer Strategies.

Corresponding Author: Douglas Faries, Eli Lilly & Company, Lilly Corporate Center, Indianapolis IN 46285, d.faries@lilly.com

## Abstract

The assumption of 'no unmeasured confounders' is a critical but unverifiable assumption required for causal inference yet quantitative sensitivity analyses to assess robustness of real-world evidence remains underutilized. The lack of use is likely in part due to complexity of implementation and often specific and restrictive data requirements required for application of each method. With the advent of sensitivity analyses methods that are broadly applicable in that they do not require identification of a specific unmeasured confounder – along with publicly available code for implementation – roadblocks toward broader use are decreasing. To spur greater application, here we present a best practice guidance to address the potential for unmeasured confounding at both the design and analysis stages, including a set of framing questions and an analytic toolbox for researchers. The questions at the design stage guide the research through steps evaluating the potential robustness of the design while encouraging gathering of additional data to reduce uncertainty due to potential confounding. At the analysis stage, the questions guide researchers to quantifying the robustness of the observed result and providing researchers with a clearer indication of the robustness of their conclusions. We demonstrate the application of the guidance using simulated data based on a real-world fibromyalgia study, applying multiple methods from our analytic toolbox for illustration purposes.

## Plain Language Summary

Analyses comparing the effectiveness or safety of interventions based on real-world (non-randomized) data are potentially biased and viewed skeptically in part due to unmeasured confounding. The assumption of 'no unmeasured confounders', that is, all variables that influence both treatment selection and outcomes are in the dataset and used in the analysis, is a requirement for producing valid real-world evidence. However, sensitivity analyses to assess this assumption is underutilized. Roadblocks to broader implementation of sensitivity analysis are decreasing with the introduction of broadly applicable methods and publicly available code. Here we propose a best practice guidance to 1) assist researchers planning a study how to measure the robustness of the design to bias from potential confounding; and 2) help researchers analyzing data to produce sensitivity analyses quantifying the robustness of their findings by understanding the strength of unmeasured confounding necessary to change study conclusions and the plausibility for such confounders to exist. A pilot study implementing the guidance on a simulated study of fibromyalgia is provided. We believe that consistent application of quantitative sensitivity analyses will provide decision makers with better information on the robustness of real-world evidence leading to greater acceptance of quality research and better patient outcomes.

## Key Points

- Our best practice guidance provides a structured approach for addressing potential unmeasured confounding at the design and analysis stages of real-world comparative research.
- Quantitative assessment of potential unmeasured confounding begins at study design: tools include directed acyclic graphs to identify confounders, benchmarking to quantify robustness, and plans for collecting information to reduce uncertainty.
- Analyses should quantify the amount of confounding necessary to change inferences, assess the plausibility of such confounders, and utilize external information to update effect estimates.
- Consistent application of quantitative sensitivity analyses will provide decision-makers with information on the robustness of RWE leading to greater acceptance of quality RWE.

## 1 Purpose

The growing availability of real-world data (RWD) sources has driven the use of real-world evidence (RWE) in the drug development process, from discovery to phase IV research and market access. This has been spurred by the 21$^{st}$ Century Cures Act and subsequent efforts considering the use of RWE to inform regulatory decisions regarding the effectiveness and safety of medical products. However, the promise of timely and credible RWE to inform regulators and healthcare decision makers is challenged by the need to address potential biases inherent in non-randomized research.

Generating credible RWE to estimate causal treatment effects requires making four key assumptions: the stable unit treatment value assumption (SUTVA), positivity, correct statistical modeling, and strong ignorability (or 'no unmeasured confounders'). The 'no unmeasured confounders' assumption is not verifiable and is a major roadblock to the acceptance of RWE for healthcare decision making[1,2]. Despite this concern, common practice simply lists the potential for bias due to unmeasured confounders as a limitation without quantitative sensitivity analyses[3,4]. Best practice guidance for RWE, produced by the International Society of Pharmacoeconomics and Outcomes Research and the International Society of Pharmaoepidemiology[3,5,6,7], emphasize the importance of addressing unmeasured confounding but do not provide a roadmap for implementation.

Several review articles[8,9,10] provide summaries of methods for addressing unmeasured confounding but are under-utilized for multiple reasons. First, many methods are complex and require custom programming for implementation. Second, many methods are only applicable in specific settings and not broadly applicable[10]. For instance, propensity score calibration requires a subsample of patients with data on the unmeasured confounder and the prior rate ratio requires data in a prior time-period. Thus, without comparisons of operating characteristics of different methods, researchers are challenged with understanding what options are available for their study.

Recently, software to implement methods that are broadly applicable (requires no knowledge of a specific unmeasured confounder or additional data) have become publicly available such as the R-packages TreatSens, Sensmakr, Evalue, and Tipr. Researchers have proposed such methods as a first step for sensitivity analyses for comparative observational research[11,12,13,14]. Zhang and colleagues[12] developed a flowchart to help researchers navigate the options and produce sensitivity analyses appropriate for their study. However, this has not undergone assessment via pilot studies and only addresses sensitivity at the analysis stage (not the design stage). Also, expanded analytical options are available since that time.

Building on the work of Zhang et al.[12], we propose a best practice guidance for both the design and analysis stages of comparative observational research, including a set of guiding questions and a toolbox of methods to help researchers quantitatively address the potential for unmeasured confounding. We illustrate the use of the guidance using simulated data based on a prospective real-world study of fibromyalgia[15] (REFLECTIONS).

## 2 Sensitivity Analysis Guidance and Toolbox for Study Design and Analyses

### 2.1 Study Design Stage

The need for planning sensitivity analyses for potential bias begins at the design stage of comparative observational research. Prior to developing a protocol, researchers should ensure the planned design and data will support a robust conclusion. Figure 1 provides a structured process to guide researchers to identify potential confounders, quantify the level of confounding that would be problematic to generating robust findings, conduct a benchmarking exercise based on the expectations, and consider options to reduce uncertainty caused by potential unmeasured confounding. To assist with implementation, a toolbox of analytic methods is included. Study objectives and designs vary, but these questions are broadly applicable, such as for non-inferiority, superiority, and predictive study aims, while the application of the toolbox may differ.

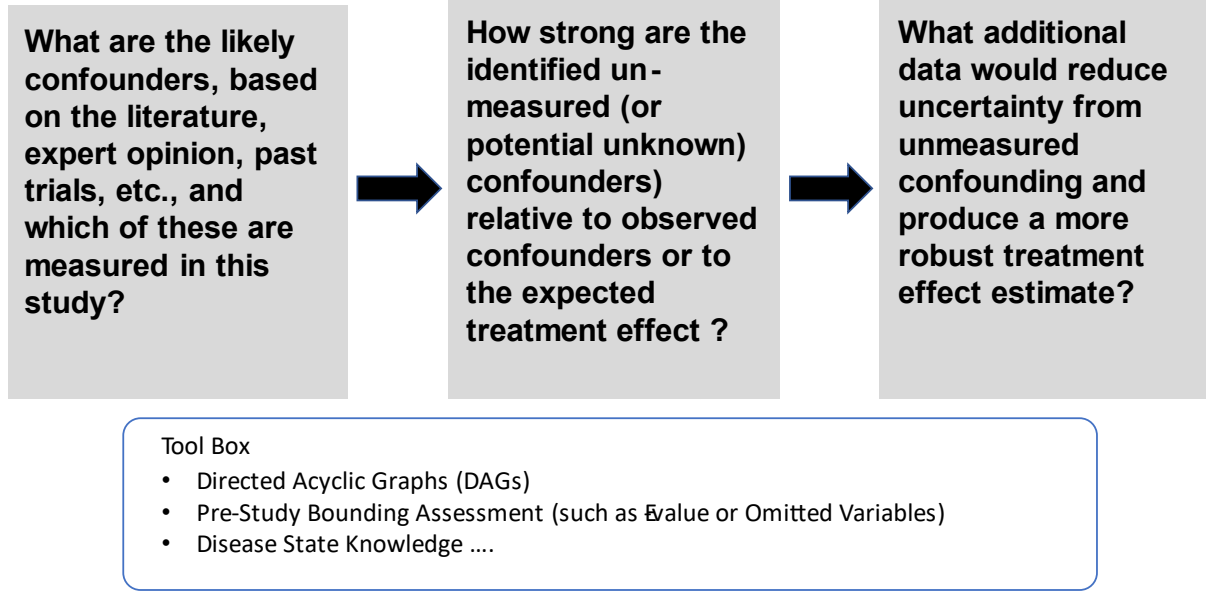

Figure 1. Pre-Study Planning: Guiding Questions & Toolbox

Girman et al.[16] recommended directed acyclic graphs (DAGs) to guide pre-study feasibility assessment in comparative real-world studies. Use of DAGs[17,18] at the design stage provides a structured approach to identify all known confounders (factors influencing both treatment selection and outcome) – measured or not – based on current evidence. Information to develop DAGs is obtained from sources such as clinical experts, prescriber surveys, literature reviews, and existing disease state data. The simple DAG in Figure 2 for a single treatment decision point illustrates that bias from unmeasured confounding is driven by two factors (after conditioning on X):

    a. Strength of association between the unmeasured confounder U and treatment Z
    b. Strength of association between the unmeasured confounder U and outcome Y.

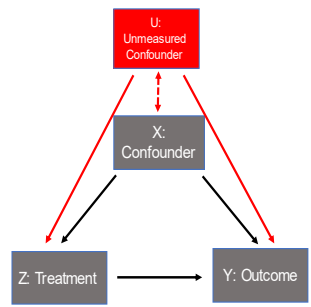

Figure 2: DAG Summarizing Measured and Unmeasured Confounding

Once potential unmeasured confounders are identified, researchers need information on the magnitude of the confounding relative to other measured confounders (for benchmarking) and the expected treatment effect. The E-value and Robustness value are statistics that quantify what level of confounding could produce the observed treatment effect when the true effect is zero. While the treatment effect is not known prior to the study, researchers have an expected treatment effect that can be used to compute a pre-study E-value or Robustness value (for both the expected treatment effect and the expected lower confidence limit).

Once bounds are computed and potential unmeasured confounders identified, benchmarking exercises compare the expected strength of confounders (based on clinical expertise or existing data) to the identified bounds of concern. Even if no specific unmeasured confounder has been identified, there is still potential bias from unknown confounders. In such cases a measured confounder serves as a proxy in the benchmarking process. That is, researchers may hypothesize that unknown confounders are not stronger than proven known confounders and then use the strongest known confounder as a conservative benchmark.

This exercise may raise or lessen concerns with unmeasured confounding. The final step is to consider the value of obtaining additional data to reduce expected uncertainty from unmeasured confounding. Options include additional data collection, such as through surveys or chart reviews, obtaining information on strength of confounding through literature reviews or disease state registries, or pre-planning intensive sensitivity analyses such as negative controls. Fang et al[19] note that while the expected effect size is fixed, the lower confidence limit varies with the sample size. Thus, if E-value demonstrates robustness to the expected observed effect but not to the lower confidence bound, increasing the sample size is one option to improve robustness at the design stage.

**2.2 Analysis Stage**

After primary data analyses are completed, sensitivity analyses address how strong unmeasured confounding would need to be to change inferences. Figure 3 provides guidance and tools for quantitative sensitivity analyses to understand the robustness of the findings to potential unmeasured confounding.

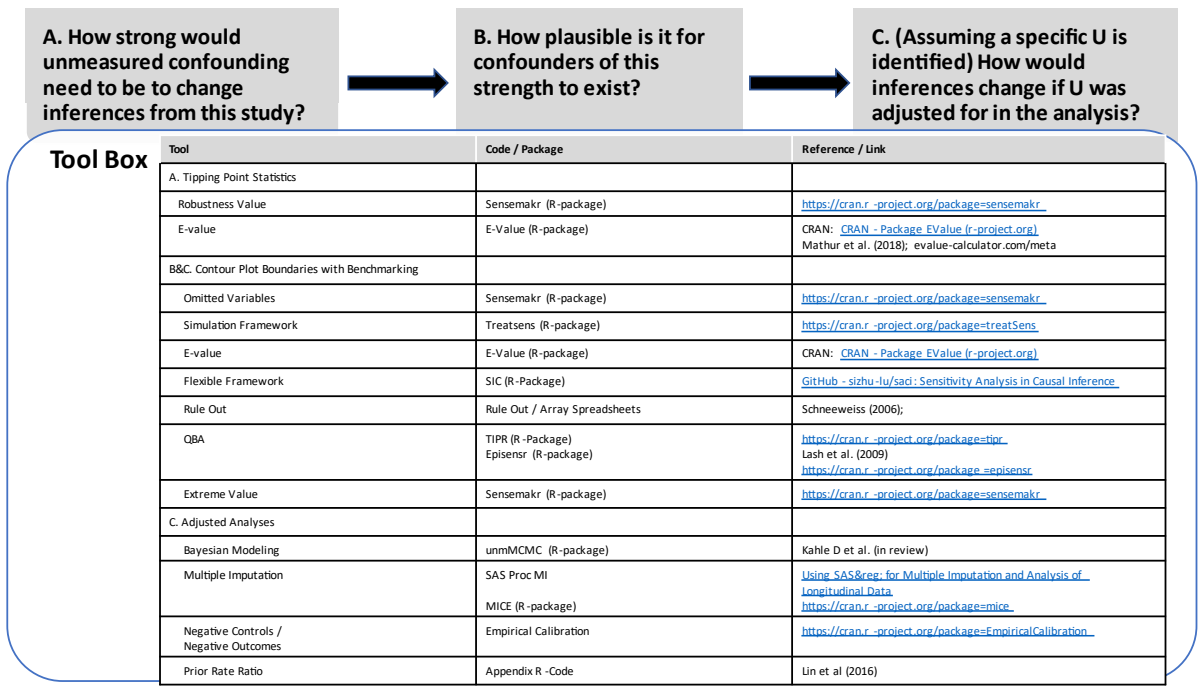

Figure 3. Analysis Stage Guiding Questions & Toolbox

There are multiple tipping point or bounding analyses available including the E-value[11], omitted variables [13], simulation framework[20,21] and array approach[22]. These methods are broadly applicable, given they require no knowledge of any specific confounder and approximations allow application for continuous, binary, and time to event outcomes. The E-value is the minimum strength of confounding necessary on a risk ratio scale that an unmeasured confounder would need to have with both the treatment and outcome to explain away a specific observed treatment effect, conditional on the observed covariates[11]. The Robustness value[13] summarizes the minimal strength of association a confounder would need with both the treatment and outcome to eliminate the observed treatment effect (or statistical significance) – quantified on a proportion of variance explained ($R^2$) scale. The simulation approach derives the conditional distribution of U given treatment, outcome, covariates and resampling generates treatment effect estimates adjusted for various strengths of U.

Contour plots are recommended for summarizing information from bounding analyses. These graphical tools display adjusted treatment effects across different pairs of assumed values for the strength of association between the unmeasured confounder and both treatment and outcome (the X and Y-axes for the contour plots). Thus, one can view the strength of confounding that would change inferences or even the direction of the treatment effect.

Once bounds are established for the strength of confounding that would change inferences, the plausibility of the existence of confounders of this strength should be considered. This requires disease-specific knowledge and clinical judgement and both benchmarking and extreme value analyses[13] can be used to address the question.

Benchmarking compares the strength of potential unmeasured confounders (based on the literature, expert opinion, or other studies) to the tipping point bounds created previously. Placing the additional information onto contour plots demonstrates both how much the treatment effect estimate may change with additional information and whether the additional information would change any inferences from the study. If the benchmarking values remain below any bounds for changing inference, then the RWE may be viewed as more robust.

If there are no known unmeasured confounders, benchmarking still proceeds using the strongest measured confounder as a reference, recognizing that such a benchmark may be conservative.

Lastly, in cases where an unmeasured confounder U is identified and additional information regarding U is available, researcher should investigate whether inferences would change if U could be adjusted for in the analysis? The set of tools for this is growing, though they are dependent on the type of information that is available on U. For instance, if the information comes from an internal subsample, then missing data methods such as multiple imputation and the Control Variable approach[23] become feasible. Bayesian methods are emerging as a useful tool as they are designed to incorporate information from various sources into the analysis. Thus, whether an internal subsample, external data, or even simply summary information from the literature or expert opinion – Bayesian methods can incorporate the information and produce an updated treatment effect estimate[24].

# 3 Pilot Application using Simulated Study data on Fibromyalgia

We piloted the best practice guidance and toolbox using our simulated REFLECTIONS fibromyalgia study. REFLECTIONS was a prospective observational study[15] that enrolled 1700 patients initiating a new treatment for fibromyalgia (new user design) and collected data longitudinally to compare 1-year pain severity outcomes (Brief Pain Inventory [BPI], null hypothesis of no treatment difference) in patients initiating opioid versus non-opioid treatments[25]. To have an example with known levels of unmeasured confounding and known true treatment effect, we generated a simulated version of the REFLECTIONS data (N = 1000) with:

- the same set of baseline covariates with the same distributions and correlations as in the actual study;
- A newly created variable 'U' which represents an unmeasured confounder;
- New treatment (Opioid or Non-Opioid) and outcome (pain severity) variables such that there was no true treatment effect and U was related to both treatment selection and outcome.

Details on the data generation process are presented in Appendix 1. To demonstrate the use of sensitivity analysis methods that leverage external data, we also generated a fibromyalgia disease registry dataset with similar covariates and outcomes as in REFLECTIONS – but only with non-opioid treated patients.

### 3.1 Design Stage

Following the principles in Figure 1, we first created a DAG to assess whether all known confounders were collected in the study (Figure 4). For simplicity, we focused on the relationships of each covariate with treatment selection and 1-year pain severity. Since REFLECTIONS collected data on all known major confounding factors, there were no known unmeasured confounders, though clinical information at any point in time is incomplete. Thus, potential for bias due to unmeasured confounding exists and the pre-study assessment continued using measured covariates as benchmarking variables.

Prior evidence suggests baseline pain severity was likely the strongest confounder. Based on linical knowledge it was considered unlikely that unmeasured confounding from unknown sources would be stronger than confounding from baseline pain severity. Thus, baseline pain severity served as a conservative benchmark for potential unknown confounding in the sensitivity analysis below.

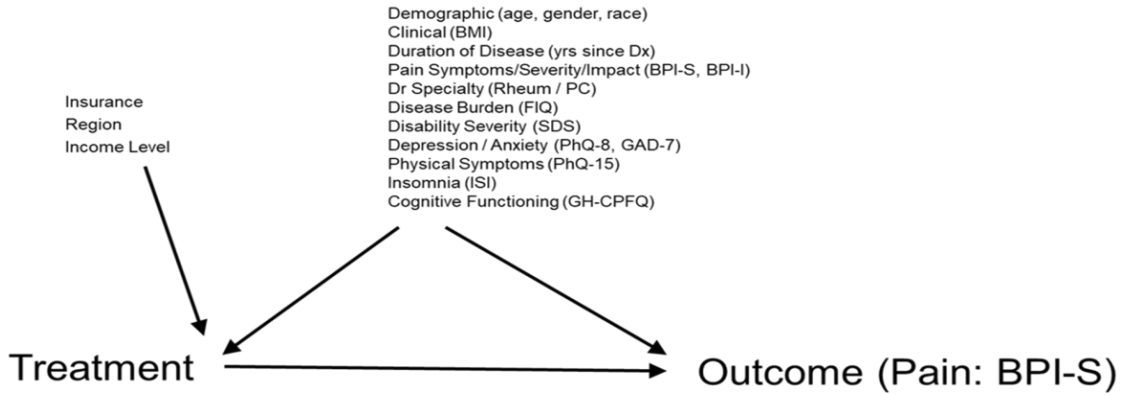

Figure 4 REFLECTIONS study DAG

To assess the robustness of the study design against unknown confounding benchmarked at the level of 'baseline pain severity', a pre-study E-value was computed. Based on a 0.25 expected treatment effect, a standard deviation of 1.2, and planned sample sizes of 200/400 per group, the E-values for the point estimate and lower confidence limit would be 1.82 and 1.31. The external registry risk ratio (RR) for baseline pain severity with pain severity outcome was 3.1. Given this was greater than 1.31, the study design was not likely to produce RWE robust against unmeasured confounding at the level of strength of baseline pain severity.

The final step of the pre-study process is to consider opportunities to collect additional information regarding potential unmeasured confounders to produce a more robust treatment effect estimate. The benchmarking calculations indicate additional information on unmeasured confounding may be beneficial for making more robust conclusions. For demonstration purposes, we assume a specific unmeasured confounder was identified and researchers made plans to obtain patient level data on the unmeasured confounder at a limited number of investigational sites (e.g. chart reviews or patient surveys). These partial data are used below to provide additional sensitivity analysis.

### 3.2 Analysis Stage

We conducted several common analyses (propensity score matching, Targeted Maximum Likelihood Estimation (TMLE), regression, and inverse probability weighting (IPW)) to estimate the effect of treatment on 1-year pain severity using only the measured baseline covariates to control for bias (ignoring the unmeasured confounder U). Results generally showed a statistically significantly greater pain reduction in Opioid treated patients with effects ranging from 0.20 to 0.47 on the BPI-Pain scale:

- ANCOVA: -0.30 (-0.55, -0.05)
- 1:1 Propensity Score Matching (ATT analysis): -0.20 (-0.49, 0.09)

- TMLE (ATE): -0.37 (-0.64, -0.10)
- AIPW (ATE): -0.47 (-0.88, -0.06).

As a 1st step to assess the robustness of the findings, we evaluated how strong unmeasured confounding must be to change inferences from the study. Though use of all methods would not be required in practice, we demonstrated 4 different tipping point analyses: the E-value, omitted variables, simulation framework, and array approach. Figure 5(a-d) displays the results. For brevity, these plots include the benchmarking values addressed later in the plausibility exercise. From Figure 5a, the E-values for the point estimate and lower confidence limit were 1.7 and 1.2. Thus, a binary unmeasured confounder associated with a RR of at least 1.7 for both treatment selection and pain severity, conditional on the measured covariates, could fully explain the observed risk ratio without any true treatment effect. Similarly, if there were an unmeasured confounder with a RR of at least 1.2 with both treatment selection and pain severity then the statistically significant finding would become non-significant.

The omitted variables and simulation framework contour plots (Figure 6b,c) display combinations of strength of confounding that could fully explain the observed result (red dashed line) or eliminate the statistical significance (blue dashed line). The strength of association is given on a partial $R^2$ scale, though the simulation framework typicaly utilizes scales based on the regression coefficients. They show that an unmeasured confounder that can explain 5% of the variance of both treatment selection and outcome could result in a loss of the observed statistical significance. Using the Sensmakr R-package, the Robustness values for the treatment effect and lower confidence limits were 0.077 and 0.012. Thus, an unmeasured confounder explaining only 1.2% of the residual variance in both the outcome and treatment models could change the treatment effect estimate enough to eliminate the statistical significance. Similarly, the array approach provides similar information using a 3-dimensional view – incorporating the prevalence of the binary unmeasured confounder as a third dimension. Results from an Extreme Value analyses[13] and causal gap analysis[26] provided similar conclusions (data not shown) and other recent broadly applicable approaches[27] are available in the toolbox as well.

Figure 5. Tipping Point Sensitivity Analysis Boundary Plots: E-value, Omitted Variables, Simulation Framework, and Array Approach.

a) E-value

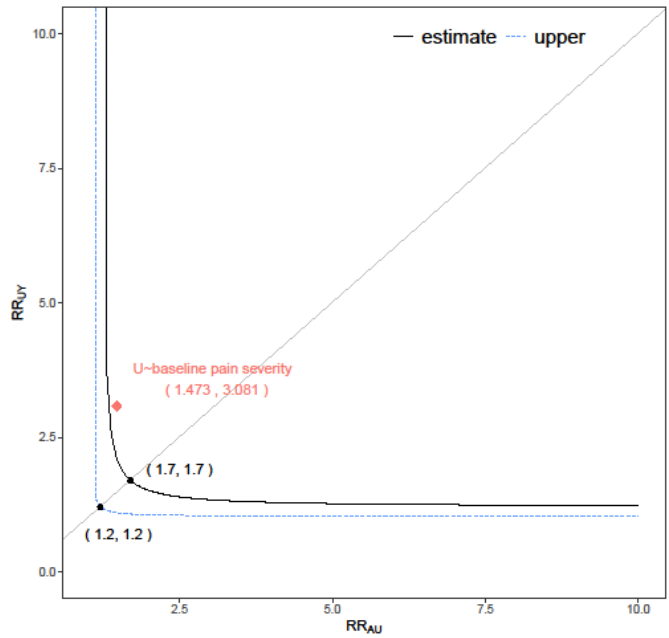

b) Omitted Variables

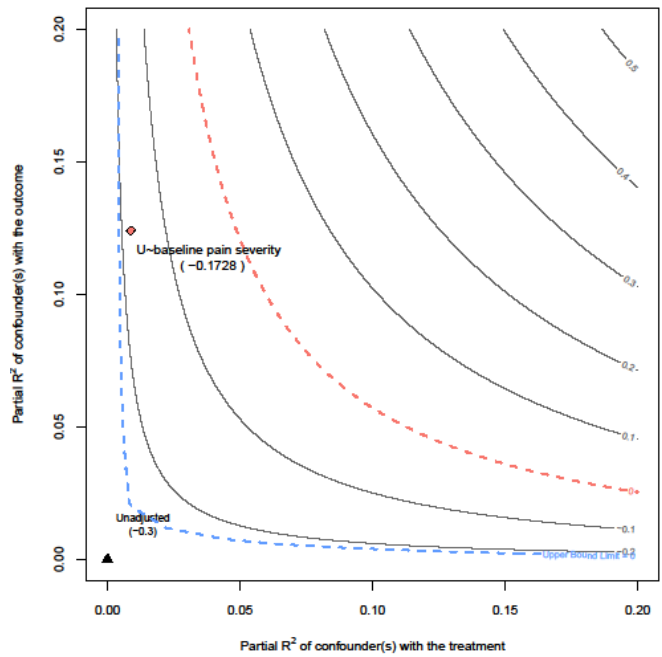

c) Simulation Framework

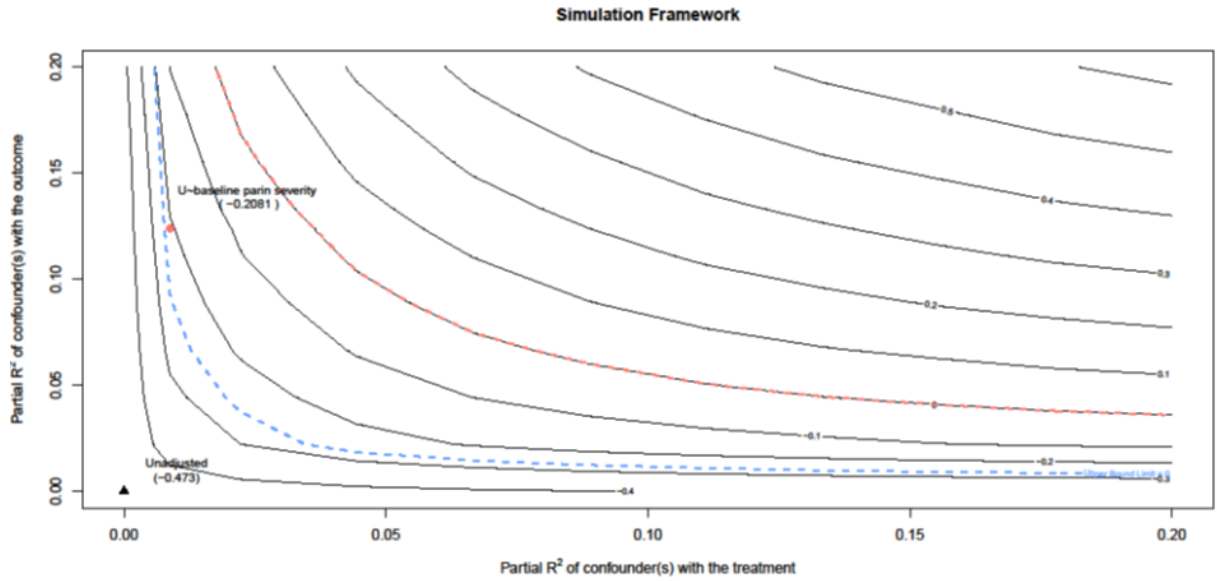

d) Array Approach

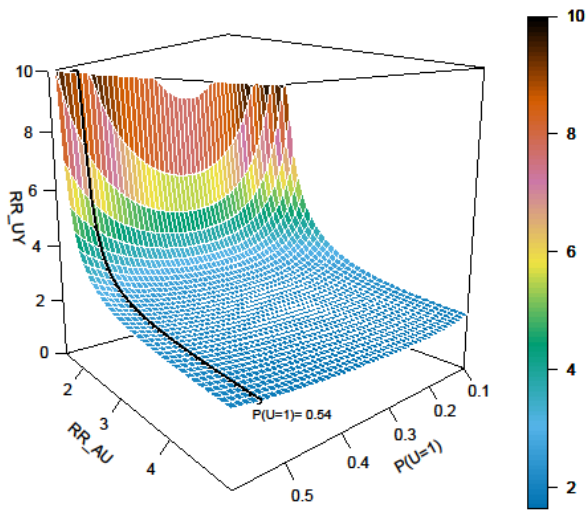

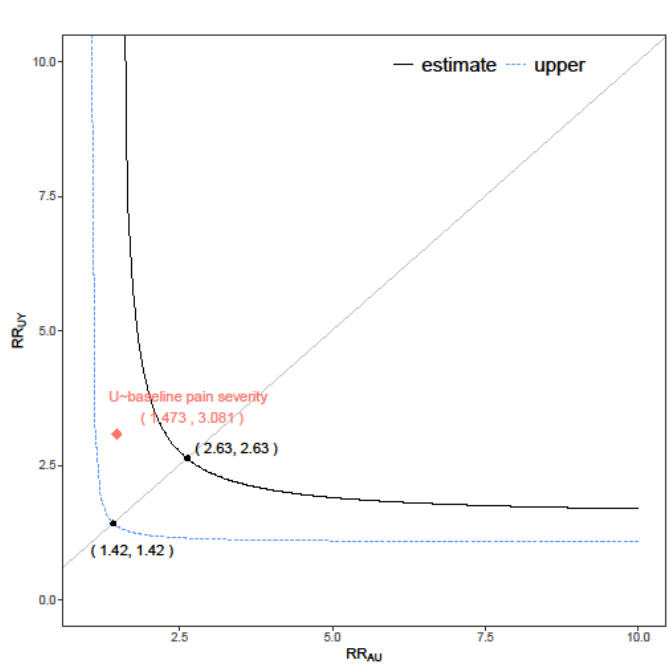

After quantifying the strength of confounding necessary to change study inferences, we then evaluated the plausibility for confounders of this strength to exist. Previously we assumed an unmeasured confounder would not exceed the level of confounding produced by our strongest measured confounder (baseline pain severity). The partial $R^2$ and RR for baseline pain severity with outcome ($R^2 = 0.12$, RR = 3.08) and treatment selection ($R^2 = 0.009$, RR = 1.47). We then updated the contour plots of Figure 5 with this benchmarking value.

Figure 5 shows the observed statistically significant result is not robust to unmeasured confounding at the strength of baseline pain severity. That is, if such an unmeasured confounder existed and could be accounted for in the analysis, the observed effect would no longer be statistically significant. Figure 5 also shows the treatment effect under any assumption regarding the strength of an unmeasured confounder. From Figure 5b, though an observed effect of -0.3 was found, the true treatment effect would be close to -0.2 accounting for an unmeasured confounder of the strength of baseline pain severity.

Given the potential lack of robustness from the initial analyses, we next consider whether incorporating additional information on the unmeasured confounder can provide greater clarity. Using information from our external registry, where the prevalence of U was 0.54, we can sharpen the benchmarking exercise. The right side of Figure 5d displays a contour plot updated with the assumption that $P(U=1) = 0.54$. While there are some differences from Figure 6a (where information on prevalence was not utilized), the same concerns with robustness remain.

Lastly, we consider whether incorporating 'internal' information on the unmeasured confounder can provide greater clarity on the robustness of our findings. Table 1 displays the results of incorporating this patient level information on U from a subset of patients through multiple imputation, the control variable approach, and Bayesian regression modeling. Multiple imputation was performed using the MICE R-package and models leveraged all baseline and outcome information. The Control Variable approach[23] first computes an initial internal estimator that adjusts for the observed confounders and U, which is consistent but not necessarily efficient. It then improves efficiency by calibrating an error-prone estimator from the internal source to the larger main data source. For the Bayesian approach, the twin regression model[10,28] was applied with relatively non-informative priors on all parameters. All three methods produced similar treatment effect estimates, ranging from -0.12 to -0.17, that were closer to the null truth than the original biased estimates. W while the data generating model had no treatment effect, the treatment effect estimate for our single random sample, when using the correct regression model with U and all other covariates in the model was -0.08. Thus, these internal methods produced estimates close to the best possible analysis from the dataset.

The internal data (partial $R^2$ values of U with treatment and outcome of 0.022 and 0.122) was also used for a more precise benchmarking exercise using the contour plots in Figure 5. Figure 5 b,c would suggest a true effect of approximately -0.10 under this level of unmeasured confounding (see Table 1, columns labelled 'Omitted Variables' and 'Simulation Framework'). Thus, the internal information successfully generated estimates closer to the truth than either the original biased analyses or using the conservative 'baseline severity' variable for benchmarking.

Table 1. Summary of Sensitivity Analyses Using Internal Information on U

| Method | Adjusted Treatment Effect Estimate | Standard Error | 95% Confidence Interval |
|---|---|---|---|
|  |  |  |  |
| Multiple Imputation | -0.15 | 0.14 | (-0.43, 0.13) |
| Control Variable Approach | -0.14 | 0.26 | (-0.65, 0.38) |
| Bayesian Modeling | -0.17 | 0.12 | (-0.41, 0.07) |
| Benchmarking |  |  |  |
|   Omitted Variables | -0.10 | 0.12 | (-0.34, 0.14) |
|   Simulation Framework | -0.09 | 0.12 | (-0.33, 0.15) |

In summary, the original analyses ignoring the unmeasured confounder suggested a statistically significantly greater reduction in pain severity for patients in the Opioid treatment group relative to the Non-Opioid group. However, quantitative sensitivity analyses suggested the statistically significant result was not robust against unmeasured confounding at the strength of baseline pain severity. Whether this would be concerning requires clinical judgement on the possibility of an unmeasured confounder of this strength. Incorporating additional information on a potential unmeasured confounder from either external or internal data sources produced much smaller treatment effect estimates close to the true value.

## Section 4 - Discussion

Comparative studies using RWE should include pre-planned and quantitative assessment of the potential impact of unmeasured confounding. As reliance on RWE by decision makers grows, so should consistent use of quality sensitivity analyses surrounding our core statistical assumptions.

Here we proposed a structured approach to guide the evaluation and correction of potential bias due to unmeasured confounding both at the stage when developing a protocol and at the data analysis stage. In each stage we present 3 questions that help researchers address the potential impact of unmeasured confounding along with a toolbox of methods and links to software for implementation. The goal at the design stage is to ensure the planned study has potential to produce a causal treatment effect estimate robust against expected levels of bias due to unmeasured confounding and stimulate planning for additional data collection. At the analysis stage, the goal is to provide quantitative assessment of the robustness of the observed finding to potential unmeasured confounding, using all available information. We believe such an approach will provide greater information regarding the robustness of RWE to medical decision makers.

We demonstrated the use of this best practice guidance with the simulated REFLECTIONS study. While commonly used analyses found a statistically significant treatment effect, quantitative sensitivity analyses suggested the results were not robust to potential unmeasured confounding at the strength of the largest measured confounder (baseline pain severity). Incorporating additional information on an identified unmeasured confounder enabled us to provide more accurate benchmarking and adjusted treatment effect estimates close to the true treatment effect.

We note that this work is not without limitations. First, we have piloted a single application; greater use of the guidance will likely bring revisions and improvements. For instance, the internal data was beneficial in our pilot example but with a smaller internal sample it may be less helpful. Second, our toolbox will change over time as additional methods and software options become available. We did not demonstrate all potentially useful methods, such as empirical calibration based on negative controls or instrumental variables analyses. Other designs such as non-inferiority studies and scenarios with a null finding in the statistical analyses have not been addressed here. Unmeasured confounding is just one of several assumptions required for causal inference and we recommend a careful evaluation of the validity of all assumptions. We focused on unmeasured confounding given its potential to cause significant bias and the under-utilization of sensitivity analyses for this application.

In summary, consistent application of quantitative sensitivity analyses will provide decision makers with better information on the robustness of RWE. This transparency will lead to greater acceptance of quality RWE and ultimately better patient outcomes. Our hope is that this work will accelerate the growing trend toward consistent and high-quality application of quantitative sensitivity analyses for unmeasured confounding.

# Ethics Statement

Not applicable.

# Conflict of Interests Information